\documentstyle[11pt,newpasp,twoside,epsf]{article}
\def\edcomment#1{\iffalse\marginpar{\raggedright\sl#1\/}\else\relax\fi}
\reversemarginpar

\begin{document}

\title{Time-resolved flickering mapping of V2051 Ophiuchi}
\author{R. Baptista and A. Bortoletto}
\affil{Dept.\ de F\'{\i}sica, UFSC, Trindade, 88040-900, Florian\'opolis,
		Brazil}

\begin{abstract}
Although flickering is one of the fundamental signatures of accretion,
it is also the most poorly understood aspect of the accretion processes. 
A promising step towards a better undestanding of flickering consists 
in using the eclipse mapping method to probe the surface distribution 
of the flickering sources.
We report on the analysis of light curves of the dwarf nova 
and strong flicker V2051 Ophiuchi with eclipse mapping techniques to 
produce the first maps of the flickering  brightness distribution 
in an accretion disc.
\end{abstract}

%\section*{Data analysis}

Time-series of CCD photometry of V2051 Oph in the B-band 
were obtained at the Laborat\'orio Nacional de Astrof\'{\i}sica, in 
southern Brazil, in 1998 and 1999, while the star was in quiescence. 
V2051~Oph was brighter in 1999 (by $\simeq 50$\%). The
flickering amplitude is clearly larger in the data of that year,
suggesting a dependency of flickering amplitude with brightness level.
Therefore, the data of the two years were analyzed separately.
We used the `ensemble' technique  (Bennie, Hilditch \& Horne 1996)
to produce lightcurves of the steady (average) and the flickering 
(residuals) components for each epoch.

The comparison of the steady light lightcurves of 1999 and 1998
indicates that the change in brightness level is mainly due to enhanced
emission from the bright spot and gas stream region in 1999, suggesting 
a long term increase in the mass transfer rate. 
The flickering curves are distinct from the corresponding steady light 
lightcurves, showing a narrower eclipse clearly displaced towards later 
phases with respect to phase zero; they also show a more prominent
orbital hump, indicating a significant (anisotropic) contribution from 
the bright spot. 

Maps of the surface brightness of the steady light and of the flickering
are obtained by applying eclipse mapping techniques (Baptista \& Steiner
1993) to these light curves (Fig.\ 1).
The eclipse maps of the steady light show the bright, compact white dwarf 
at disc centre plus a faint (and slightly asymmetric in 1998) disc 
typical of quiescent dwarf novae; 
the disc is larger and the bright spot emission at the disc rim is 
stronger in 1999, underscoring the suggestion that the mass transfer rate 
was higher than in 1998.
\begin{figure}[th]	% FIGURA 4
\plotfiddle{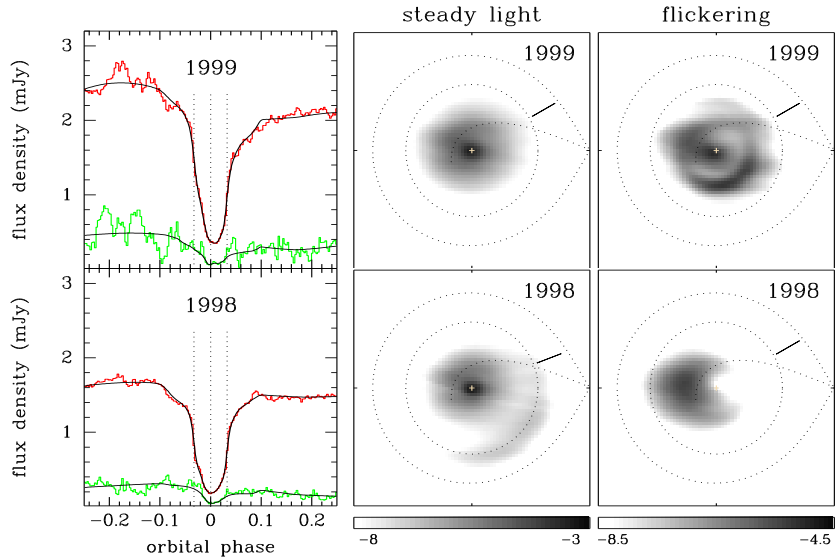}{6.5cm}{-90}{48}{48}{-190}{242}
\caption{ Left: the steady light and flickering curves (stepped curves) and
  the eclipse mapping fits (solid lines). The flickering curves were scaled 
  by a factor of 2 for a better visualization.
  Vertical dotted lines mark the ingress/egress phases of the white dwarf 
  and mid-eclipse. 
  Middle: eclipse maps of the steady light in a logarithmic greyscale. 
  Right: flickering maps in a logarithmic greyscale. 
  Dotted lines depict the primary Roche lobe, the disc radius, 
  and the gas stream trajectory. }
\end{figure}
The flickering maps are noticeably different from the corresponding maps 
of the steady light. The 1998 flickering map is dominated by emission 
from the disc side farther away from the secondary star (i.e.
the 'back' side of the disc) and from the bright spot at disc rim. 
The flickering distribution changes significantly from 1998 to 1999, 
probably in response to the overall increase in brightness of the system. 
The 1999 flickering map shows enhanced emission along the gas stream 
trajectory as well as a (reduced) contribution from the back side of 
the disc. A contribution from the inner disc regions (absent in 1998) 
is also visible.

The changes in the spatial distribution of the flickering sources 
suggest that the flickering in V2051~Oph is associated to inhomogeneities 
in the mass transfer from the secondary star.
Our interpretation of the flickering maps invokes the `magnetic
propeller' model of Horne (1999).
In a lower accretion regime (e.g., 1998), the incomming blobs 
of gas extract energy and angular momentum 
from the magnetic field of the inner disc regions (or the WD)
to be flinged out of the binary towards the back side of the disc,
producing the observed flickering distribution as they collide with 
each other and/or with the disc material.
When the mass accretion rate increases (e.g., in 1999), the gas plasma
blobs have enough specific kinetic energy to start being accreted by 
loosing angular momentum in their interaction with the inner disc (or WD)
magnetic field, sweeping around the disc roughly along the ballistic 
stream trajectory to produce the observed flickering distribution until 
they are destroyed by the action of the disc viscosity. 

% \acknowledgments
% CNPq and PRONEX are gratefully acknowledged for financial support 
% through grants no. 300\,354/96-7 and FAURGS/FINEP 7697.1003.00.

\end{document}